\documentclass[
reprint,
superscriptaddress,
amsmath,amssymb,
aps,
prb,
eqsecnum
]{revtex4-1}
\usepackage{graphicx}
\usepackage{dcolumn}
\usepackage{bm}
\usepackage{braket}
\usepackage{comment}

\makeatletter
\let\MYcaption\@makecaption
\makeatother

\usepackage[subrefformat=parens]{subcaption}
\makeatletter
\let\@makecaption\MYcaption
\makeatother

\usepackage[
colorlinks=true,
urlcolor=blue,
citecolor=blue,
linkcolor=blue,
hyperfootnotes=false]{hyperref}

\usepackage[dvipsnames]{xcolor}

\allowdisplaybreaks

\begin{document}

\title{Gapless symmetry-protected topological phase of quantum antiferromagnets \\ on anisotropic triangular strip}

\author{Yuichiro Hidaka}
\affiliation{
Institute for Solid State Physics, University of Tokyo, Kashiwa 277-8581, Japan }
\author{Shunsuke C. Furuya}
\affiliation{Department of Physics, Ibaraki University, Mito, Ibaraki 310-8512, Japan}
\affiliation{Department of Basic Science, University of Tokyo, Meguro, Tokyo 153-8902, Japan}
\author{Atsushi Ueda}
\affiliation{
Institute for Solid State Physics, University of Tokyo, Kashiwa 277-8581, Japan }
\author{Yasuhiro Tada}
\affiliation{
Quantum Matter Program, Graduate School of Advanced Science and Engineering,
Hiroshima University, Higashihiroshima, Hiroshima 739-8530, Japan }
\affiliation{
Institute for Solid State Physics, University of Tokyo, Kashiwa 277-8581, Japan }

\date{\today}
\begin{abstract}
We study a three-leg spin-1/2 ladder with geometrically frustrated interleg interactions.
We call this model an anisotropic triangular-strip (ATS) model.
We numerically and field-theoretically show that its ground state belongs to a gapless symmetry-protected topological (SPT) phase.
The numerical approach is based on density-matrix renormalization group analyses of the entanglement entropy and the entanglement spectrum.
Whereas the entanglement entropy exhibits a critical behavior, the entanglement spectrum is nontrivially degenerate.
These entanglement properties imply that the ground state is a gapless topological phase.
We investigate the ATS model using a quantum field theory to support the numerical findings.
When the frustrated interchain interaction is deemed a perturbation acting on the three spin chains, the frustrated interchain interaction almost isolates the second chain from the other two chains.
However, at the same time, the second chain mediates a ferromagnetic interaction between the first and third chains.
Therefore, the ground state of the ATS model is a gapless Tomonaga-Luttinger liquid weakly coupled to a spin-1 Haldane chain with irrelevant interactions.
Last but not least, we show that the gapless SPT phase of the ATS model is a symmetry-protected critical phase.
We point out that the symmetry protection of criticality is essential in characterization of the gapless SPT phase.
\end{abstract}

\maketitle

\section{Introduction}

The spin-1 Haldane phase~\cite{haldane_1983a,haldane_1983b,Affleck1989_review,pollmann2010entanglement,PhysRevB.85.075125,PhysRevLett.66.798} is regarded as one of the best-known examples of symmetry-protected topological (SPT) phases~\cite{PhysRevB.83.035107,PhysRevB.87.155114,pollmann2010entanglement,PhysRevB.85.075125,chen2014symmetry}. 
The SPT phase is a gapped symmetric phase that is accompanied by no local order parameter but is still distinct from the trivial phases.
What characterizes the SPT phase is a nontrivial short-range entanglement robust to any local disturbance under symmetries.
The ground state in the spin-1 Haldane phase exhibits a characteristic entanglement spectrum where every eigenvalue is even-fold degenerate under symmetries~\cite{pollmann2010entanglement,PhysRevB.85.075125}.
This nontrivial degeneracy distinguishes the spin-1 Haldane phase from the trivial phases.
The degeneracy is protected when at least one of the following three symmetries is present: (i) the $D_2\cong \mathbb Z_2\times\mathbb Z_2$ spin-rotation symmetry, (ii) the time-reversal symmetry, and (iii) the bond-centered inversion symmetry~\cite{pollmann2010entanglement,PhysRevB.85.075125}. 

Recently, a gapless analog of the SPT phase, called a gapless SPT phase, has drawn attention~\cite{PhysRevX.7.041048,PhysRevB.97.165114,PhysRevB.104.075132,PhysRevB.103.L100207,PhysRevX.11.041059,PhysRevB.101.115131,wang2020_gSPT,li2022_gSPT,kestner2011gaplesshaldane,Fidkowski2011Majoranananowire,Ruhman2012elongateddipolar,Keselman2015gaplessSPTfermion}.
While gapped SPT phases are well understood, characterization of gapless SPT phases is underway because of their nontriviality exemplified by the coexistence of gapless bulk and edge modes.
The gapped SPT phase hosts gapped excitations in bulk and gapless excitations on edges.
The existence of the finite excitation gap in bulk partly assures the stability of the edge mode.
On the other hand, the gapless SPT phase has gapless excitations both in bulk and edges.
It is highly nontrivial how these gapless modes in bulk and on edges stably coexist in the gapless SPT phase.

Scaffidi \textit{et al.} constructed a gapless SPT state by first preparing the gapped $\mathbb Z_2\times \mathbb Z_2$ SPT state and then making it gapless~\cite{PhysRevX.7.041048,PhysRevB.97.165114}.
Their argument is based on the concept of the decorated domain wall.
This gapped $\mathbb Z_2\times \mathbb Z_2$ SPT state is closely related to the spin-1 Haldane state.
The Hamiltonian of Ref.~\cite{PhysRevX.7.041048} has a minimal structure with the essence of the spin-1 Haldane state as the $\mathbb Z_2\times\mathbb Z_2$ SPT state. However, their Hamiltonian contains three-spin interactions challenging for experimental realizations.
Originally, the spin-1 Haldane phase attracted broad attention for the simple and experimentally feasible parent Hamiltonian~\cite{Hagiwara1990_edge}.
Therefore, it will be worth pursuing an experimentally feasible antiferromagnetic model as a gapless-SPT counterpart to the spin-1 Heisenberg antiferromagnetic chain.

This paper shows that a simple spin-$1/2$ three-leg spin ladder with geometrically frustrated interchain interactions (Fig.~\ref{fig:ats_model}) exhibits a gapless SPT phase.
We call this model an anisotropic triangular-strip (ATS) model. This model is relevant to real compounds such as Cu$_3$(OH)$_4$MO$_4$ for $\mathrm{M=S, Mo}$~\cite{Vilminot2003_antlerite,Okubo2009_antlerite,Fujii2009_antlerite,Hara2011_antlerite,Fujii2013_antlerite,Fujii2015_szenicsite,Lebernegg2017_szenicsite}. In the first part of the paper, we show that the ground state of the ATS model behaves as a gapless Tomonaga-Luttinger liquid (TLL)~\cite{giamarchi_book} with a topologically degenerate entanglement spectrum by using the density matrix renormalization group (DMRG) method.
Second, we give a firm theoretical foundation to the numerical findings~\cite{gogolin_book}.
Last but not least, we show that an inversion symmetry protects the criticality and the topological degeneracy of the entanglement spectrum.
In this sense, our gapless SPT phase qualifies as a symmetry-protected critical phase~\cite{furuya2017spc}.

This paper is organized as follows: In Sec.~\ref{sec:model}, we will introduce the ATS XXZ model and explain its properties. In Sec.~\ref{sec:numericalresults}, we will show the numerical calculation results of the model. Then, in Sec.~\ref{sec:Effective field theory},
we discuss the gapless SPT phase of the model by a field-theoretical approach. Finally, in Sec.~\ref{sec:symmetryprotection}, we discuss the symmetries which protect the gapless SPT phase.

\section{Model}\label{sec:model}
\begin{figure}[t]
 \begin{minipage}[b]{\linewidth}
  \centering
  \includegraphics[keepaspectratio, width=8cm]{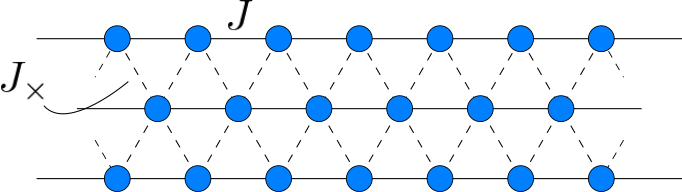}
  \subcaption{Spin-1/2 anisotropic triangular strip XXZ model.}\label{fig:ats_model}
 \end{minipage}\\\vspace{1pc}
 \begin{minipage}[b]{\linewidth}
  \centering
  \includegraphics[keepaspectratio, width=8cm]{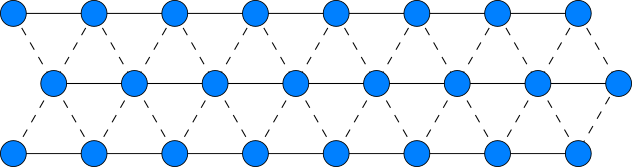}
  \subcaption{Finite-size cluster with V-shaped edges.}\label{yabane}
 \end{minipage}\\\vspace{1pc}
 \begin{minipage}[b]{\linewidth}
  \centering
  \includegraphics[keepaspectratio, width=8cm]{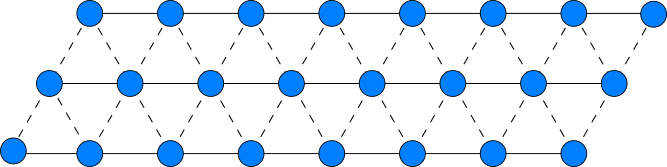}
  \subcaption{Finite-size cluster with parallel edges.}\label{parallelogram}
 \end{minipage}
 \caption{Spin-1/2 ATS XXZ model with intrachain interaction $J$ and frustrated interchain interaction $J_{\times}$. (a) Infinite chain, (b) finite-size cluster with V-shaped edges and (c) finite-size cluster with  parallel edges.}\label{obc}
\end{figure}

The ATS model has the following Hamiltonian on a three-leg ladder:
\begin{align}
H=&J\sum_{i=1}^L\sum_{n=1}^3\vec{S}_{i,n}\cdot_{\Delta}\vec{S}_{i+1,n}\nonumber\\
&+J_{\times}\sum_{n=1,3}\sum_{i=1}^L\vec{S}_{i,2}\cdot_{\Delta}(\vec{S}_{i,n}+\vec{S}_{i+1,n}),
\label{H_ATS}
\end{align}
where $\vec{S}_{i,n}$ denotes the $S=1/2$ spin operator at the $i$th site along the leg on the $n$th leg (see Fig.~\ref{fig:ats_model}), and $\vec{S}_{i,m}\cdot_{\Delta}\vec{S}_{j,n}$ denotes the XXZ interaction;
\begin{align}
\vec{S}_{i,m}\cdot_{\Delta}\vec{S}_{j,n}=\frac12(S_{i,m}^+S_{j,n}^-+S_{i,m}^-S_{j,n}^+)+\Delta S_{i,m}^zS_{j,n}^z.
\end{align}
We consider antiferromagnetic exchange couplings $J>0$ and $J_\times>0$.
We also limit ourselves to a situation $0<1-\Delta \ll 1$ of the weakly easy-plane anisotropy.
We denote the system length as $L$ and the total number of spins as $3L$.

Before giving detailed numerical and field-theoretical discussions, let us briefly explain why we expect the gapless SPT phase in this ATS model \eqref{H_ATS}.
The key point is that geometrically frustrated interactions [the second line of Eq.~\eqref{H_ATS}] couple the second (middle) spin-$1/2$ XXZ chains with the first (upper) and the third (lower) chains (Fig.~\ref{fig:ats_model}).
Let us regard $J_\times$ as a perturbation to three decoupled spin chains.
We can expect that a second-order process will yield a direct ferromagnetic interchain interaction between the first and third chains of $\sim -J_\times^2/J$ (see Sec.~\ref{sec:Effective field theory}).
On the other hand, the geometrical frustration of the interchain interactions suppresses the antiferromagnetic correlation between the second chain and the other two chains.
Hence, we expect the spin-1/2 ATS model to effectively turn into a set of an almost isolated spin-1/2 XXZ chain and a two-leg spin-1/2 ladder.
The former consists of the second chain, and the latter the first and third chains.
This almost decoupled spin-1/2 XXZ chain behaves as the gapless TLL at low energies. 
On the other hand, the first and third chains effectively form a spin-1/2 two-leg ladder with the ferromagnetic interchain interaction. 

It is well known that this spin ladder can have the spin-1 Haldane state as its ground state~\cite{hijii2005_xxz_ladder}.
Therefore, we can naively expect the ground state that may be approximated as a tensor product state of the TLL from the second chain and the SPT (spin-1 Haldane) state from the other chains.
In what follows, we demonstrate that the ground state of the ATS model \eqref{H_ATS} is essentially the tensor product state that we guessed here.
Moreover, we show that even when the ground state is the product state of the gapless state and the gapped SPT state, their symmetry protection differs from that of the gapped SPT phase because the Hamiltonian contents are \emph{not} decoupled.
We will come back to this point later in Sec.~\ref{sec:symmetryprotection}.

\section{Numerical results}\label{sec:numericalresults}

This section presents the numerical results of the ATS model \eqref{H_ATS}.
Throughout this section, we set $J=1$, $J_{\times}=0.5$, and $\Delta=0.8$.
We performed the finite-size density-matrix renormalization group (DMRG) calculation and infinite-size DMRG (iDMRG) calculation to obtain the ground state and investigate its properties. 
We used the ITensor library \cite{itensor} for the finite-size DMRG calculations,
where we used the bond dimension $m\le 2400$ and kept the truncation error up to $10^{-6}$. 
We confirmed the convergence of the ground energy and entanglement entropy at the center bond within the sweep count.
On the other hand, we used the TeNPy library \cite{tenpy} for the iDMRG calculations, where we used the bond dimension $m\le 1200$.

The finite-size DMRG calculations were performed under the open boundary condition.
We consider two kinds of finite-size clusters, one with V-shaped edges (Fig.~\ref{yabane}) and the other with parallel edges (Fig.~\ref{parallelogram}).
One might expect that they differ only in the shape of the left and right edges, which would be insignificant to bulk quantities.
This expectation is mostly the case.
However, the difference in the cluster shapes becomes vital in, for example, entanglement properties because of symmetries in the corresponding Hamiltonians.
With the V-shaped edges, the system breaks the inversion symmetry, while the system with the parallel edges does not.
We will specify the shape of the system when we refer to the finite-size DMRG results.

SPT phases are characterized by no local order parameters.
In some gapped SPT phases, nonlocal order parameters are still available.
For example, a string order parameter allows us to distinguish the spin-1 Haldane phase from the trivial phases~\cite{den1989preroughening,nakamura2002_string}. 
A possible string order parameter in our model is discussed in Appendix~\ref{app:string}.
In general, however, we cannot \textit{a priori} expect the existence of the nonlocal order parameter that characterizes the SPT phase.
Instead of relying on an order parameter, one can characterize the SPT phase with the entanglement entropy and the entanglement spectrum under symmetries~\cite{pollmann2010entanglement,PhysRevB.85.075125}.
For example, the even-fold degenerate entanglement spectrum of the spin-1 Haldane state enables one to distinguish it from the topologically trivial phase. 

We employ the same strategy to characterize the gapless SPT phase.
Namely, in order to conclude that the ATS model \eqref{H_ATS} has the gapless SPT phase, we confirm the following two properties of the system. 
First, we check that the ground state is gapless by investigating the ground state's entanglement entropy.
Next, we check the even-fold degeneracy in the entanglement spectrum.
Furthermore, we will show that these properties are protected by symmetries later in Sec.~\ref{sec:symmetryprotection}.

\subsection{Entanglement entropy and central charge}\label{sec:eecentral}

Let us discuss the entanglement entropy and central charge.
We divide the whole system into two subsystems $A$ and $B$. 
We define the reduced density matrix of the subsystem $A$, by using the ground state $\Ket{\psi}$ of the whole system, as
\begin{align}\label{eq:reduceddensitymatrix}
\rho_A=\mathrm{Tr}_B\Ket{\psi}\Bra{\psi},
\end{align}
where $\mathrm{Tr}_B$ denotes the trace over the other subsystem $B$.
The entanglement entropy of the subsystem $A$, $S(A)$, is defined as the von-Neumann entropy of the reduced density matrix $\rho_A$~\cite{PhysRevLett.90.227902,latorre2003ground},
\begin{align}
    S(A)=\mathrm{Tr}_A\left[-\rho_A\ln\rho_A\right]=-\sum_{i=1,2,\cdots}\lambda_i\ln\lambda_i,
    \label{S_A}
\end{align}
where $\{\lambda_1, \lambda_2, \cdots\}$ are the eigenvalues of the matrix $\rho_A$. 
The entanglement entropy can characterize how strongly the subsystem $A$ is entangled to the rest of the system, $B$. 
For example, if the state is the product state of subsystem $A$ and $B$, $\Ket{A}\otimes\Ket{B}$, the entanglement entropy \eqref{S_A} vanishes because $\{\lambda_1,\lambda_2,\cdots \}=\{1\}$.

Throughout this paper, we deal with the one-dimensional model. 
We can thus take the subsystem $A$ as a one-dimensional system with the length $l$ from the left edge.
Note that this subsystem $A$ contains the $3l$ sites.
Therefore, we can represent the entanglement entropy as $S(l)$. 
It is reported that if the system obeys a conformal field theory (CFT) with the central charge $c>0$, the entanglement entropy is given by the following Calabrese-Cardy formula~\cite{calabrese2004entanglement,igloi2008finite}:
\begin{align}\label{eq:CalabreseCardy}
    S(l)=\frac{c}{6}\ln\left[\frac{2L}{\pi}\sin\biggl({\frac{\pi l}{L}}\biggr)\right]+c'_1+\ln g,
\end{align}
where $c'_1$ is a constant, and the $\ln g$ term represents boundary effects proportional to $\ln g\sim (-1)^l/\left(\frac{L}{\pi}\sin\left(\frac{\pi l}{L}\right)\right)$~\cite{PhysRevLett.96.100603}. 
The formula \eqref{eq:CalabreseCardy} is generically valid in a one-dimensional system with the open boundary condition.

\begin{figure}[t]
\centering
\includegraphics[width=8cm]{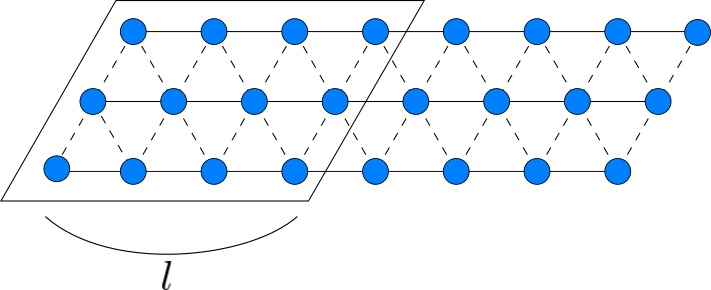}
\caption{Subsystem of ATS XXZ model used for calculations of entanglement entropy and central charge.}
\label{fig:subsystem_parallelogram}
\end{figure}

\begin{figure}[t]
\centering
\includegraphics[width=9cm]{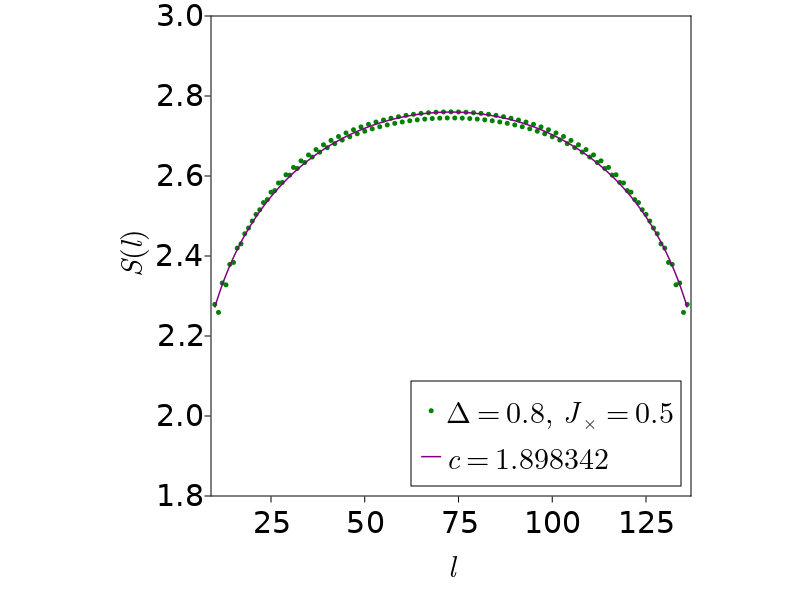}
\caption{Entanglement entropy for $\Delta=0.8$, $J_{\times}=0.5$, with $L=146$. The horizontal axis is the length of the subspace $l$ (cf. Fig.~\ref{fig:subsystem_parallelogram}), and the vertical axis is the values of the entanglement entropy. 
The solid line represents the fitting result by the Calabrese-Cardy formula \eqref{eq:CalabreseCardy}, where we have subtracted the edge term for better visibility.}
\label{fig:entent}
\end{figure}

In the ATS XXZ model, we take the subsystem as shown in Fig.~\ref{fig:subsystem_parallelogram}.
For the calculations of the entanglement entropy, we adopt the finite-size system with the parallel edges (Fig.~\ref{parallelogram}) because the system is then symmetric under an inversion, 
\begin{align}
    \bm S_{j,n}\to \bm S_{L+1-j,4-n}. 
    \label{inversion}
\end{align}
We calculate the entanglement entropy for each value of $l$ and fit the data using the Calabrese-Cardy formula \eqref{eq:CalabreseCardy}. 
As a typical example of the calculation results,
Fig.~\ref{fig:entent} shows the entanglement entropy of the ATS XXZ model with $\Delta=0.8$, $J_{\times}=0.5$, and the system length $L=146$. In the data shown in Fig.~\ref{fig:entent}, we subtracted the edge term from the numerically-calculated entanglement entropy for better visibility. As a result, the numerical data in Fig.~\ref{fig:entent} is well fitted with Eq.~\eqref{eq:CalabreseCardy} with the central charge $c\approx1.90$.
This agreement with the Calabrese-Cardy formula \eqref{eq:CalabreseCardy} means that the system is indeed gapless and described by a CFT at low energies.

\begin{figure}[t]
\centering
\includegraphics[width=8cm]{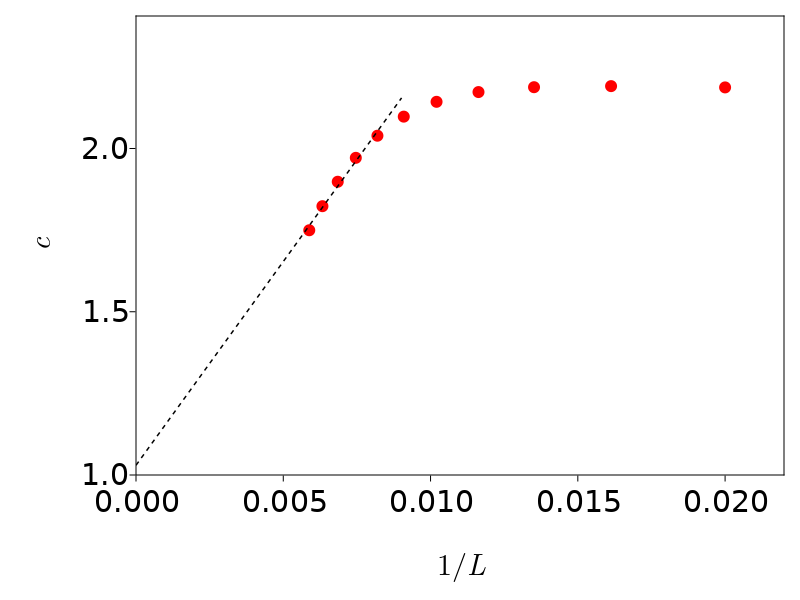}
\caption{Central charge $c$ derived from the entanglement entropy. The horizontal axis is $1/L$, and the vertical axis is the value of the central charges. The calculated value of central charge approaches $c\approx 1$ as the system length $L$ increases.}
\label{fig:central}
\end{figure}

To evaluate the central charge in the infinite-size limit $L\to+\infty$, we calculated the system-size dependence of the central charge.
Figure~\ref{fig:central} shows the numerically evaluated  central charges (the filled circles) for various system lengths $L$.
Each value of the central charge is derived in the same way as written above. 
The central charge exhibits an interesting system-size dependence.
When the system length $L$ is short, the central charge looks almost constant.
Still, it seems that the central charge with long enough $L$ decreases as $L$ increases.
Here, we simply fit the data points for the five largest system lengths with a straight dashed line in Fig.~\ref{fig:central}. 
The fitted line implies that the central charge $c$ is close to one, $c= 1.02\pm0.06$, in the $L\to+\infty$ limit, and correspondingly the ground state is a $c=1$ TLL. We find the similar value $c\sim 0.93$ when fitting the data by $1/(L\ln L)$, considering the logarithmic correction in the vicinity of the $\mathrm{SU(2)}$ point.

Note that we would overestimate the central charge $c$ in the $L\to+\infty$ limit if we used the data with a small $L$ only.
This behavior of the central charge will be attributed to the possible presence of nontrivial edge states in the gapless SPT state. 
As we show in the next subsection, the ground state of this system belongs to the gapless SPT phase whose topological property is akin to the spin-1 Haldane state.
As is well known, the spin-1 Haldane state is accompanied by a spin-1/2 state on each edge of the system.
In bulk, magnetic excitations cost a finite excitation energy whose minimum value is called the Haldane gap~\cite{TodoKato2001_HaldaneGap}.
The presence of the bulk gap makes the edge state well-localized around the edges.
On the other hand, the bulk gap is infinitesimal in the gapless SPT phase.
In particular, the bulk spin gap is infinitesimal in our case.
Therefore, the edge spins can be extended deep inside the bulk~\cite{PhysRevX.7.041048}.
When the system size is too small, we would overcount the number of gapless bulk modes.
Since the central charge reflects the number of gapless modes, it will be overestimated for small systems.
Note that such an effect was also reported by Nataf \textit{et al.}~\cite{PhysRevB.104.L180411} in a critical $\mathrm{SU}(3)$ spin chain.

\subsection{Entanglement spectrum}\label{sec:es}
Now that we confirmed that the criticality of the ground state, we investigate topological properties of the system based on the entanglement spectrum.
To calculate the entanglement spectrum, we divide the whole system into 2 subsystems $A$ and $B$, as in Sec. \ref{sec:eecentral}. The reduced density matrix of the subsystem $A$ has been defined in Eq.(\ref{eq:reduceddensitymatrix}). By using the reduced density matrix $\rho_A$, we can define the entanglement spectrum $\{\mu_i\}_{i=1,2,\cdots}$ as $\mu_i=-\ln\lambda_i$, where $\lambda_i$ are the eigenvalues of the reduced density matrix $\rho_A$~\cite{li2008entanglement,pollmann2010entanglement}.

Let us recall the relation of $\mu_i$ to the ground state $\Ket{\psi}$ of the system.
In general, $\Ket{\psi}$ is written as
\begin{align}
\Ket{\psi}=\sum_{i,j}M_{ij}\Ket{i}_A\Ket{j}_B,
\end{align}
where $M_{ij}$ is an $N_A\times N_B$ matrix, where $N_A$ is the dimension of the Hibert space of the subsystem $A$, and $N_B$ is that of the subsystem $B$. By using the singular value decomposition, we can diagonalize the matrix $M_{ij}$ and obtain;
\begin{align}
\Ket{\psi}=\sum_{I=1,2,\cdots}\Lambda_{I}\Ket{I}_A\Ket{I}_B,
\end{align}
with $\Lambda_I \geq 0$.
Then, the entanglement spectrum $\{\mu_i\}_i$ is defined as 
\begin{align}
    \Lambda_i^2 = e^{-\mu_i}.
\end{align}

It was reported that the spin-1 Haldane state shows the even-fold degeneracy of the entanglement spectrum~\cite{pollmann2010entanglement}.
Hence, we can naively expect the same even-fold degenerate entanglement spectrum in our gapless SPT phase, as briefly explained below.
Generally, for a simple tensor product of a gapped SPT ground state $\ket{\psi_1}$ and a gapless ground state $\ket{\psi_2}$,
the entanglement spectrum is calculated as $\{\mu_{1,i}+\mu_{2,j}\}_{i,j}$, where $\{\mu_{1,i}\}$ is the entanglement spectrum of $\ket{\psi_1}$ and $\{\mu_{2,i}\}$ is that of $\ket{\psi_2}$. 
Since every $\mu_{1,i}$ has its partner $\mu_{1,i'}$ so that $\mu_{1,i}=\mu_{1,i'}$ with $i\not=i'$, the entanglement spectrum $\{\mu_{1,i}+\mu_{2,j}\}_{i,j}$ of the tensor-product state is even-fold degenerate.
Therefore, we can expect that an essentially same even-fold degenerate entanglement spectrum is found in our gapless SPT state, if we regard it as the effective product state.

\begin{figure}[t]
\centering
\includegraphics[width=8cm]{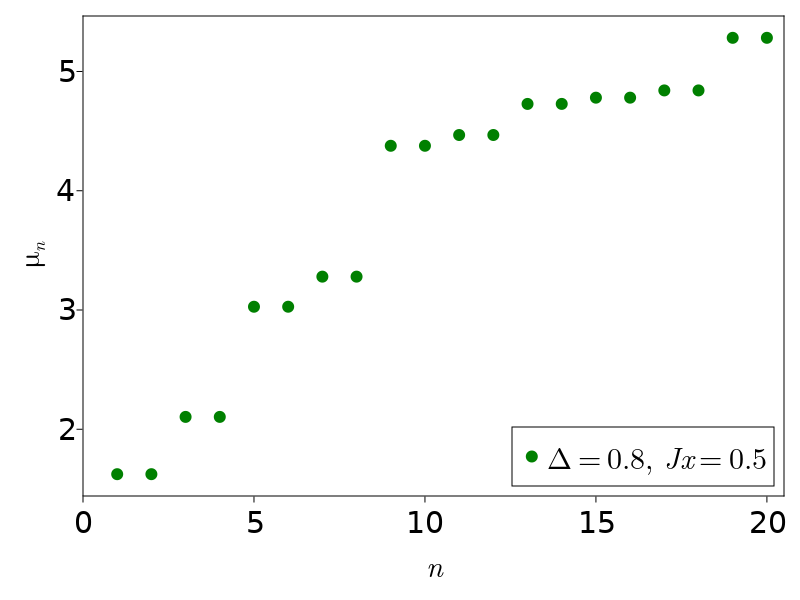}
\caption{The entanglement spectrum of ATS XXZ model for $\Delta=0.8$, $J_{\times}=0.5$, $L=100$ at center bond. This result is calculated with the finite DMRG method, and we took the lowest-energy state in the sector where the total magnetization is $1$.}
\label{fig:entspec_finite_parallelo}
\end{figure}

Figure~\ref{fig:entspec_finite_parallelo} shows a DMRG result of the entanglement spectrum of the ATS XXZ model with $\Delta=0.8$, $J_{\times}=0.5$ with the system length $L=100$.
We use the system with the parallel edges (Fig.~\ref{parallelogram}) and impose the open boundary condition.
To calculate the entanglement spectrum, we cut the system at a center bond ($l=L/2$) invariant under the inversion \eqref{inversion}.
As Fig.~\ref{fig:entspec_finite_parallelo} shows, we observe the even-fold degeneracy for every eigenvalue $\mu_i$.
Note that the inversion symmetry is critical to protect the even-fold degeneracy.
In the entanglement spectrum of the model with the V-shaped edges, we did not observe clear even-fold degeneracy because the model breaks the inversion symmetry in the finite-size system, and the calculation results suffer from rather significant finite-size effects.

\begin{figure}[t]
\centering
\includegraphics[width=8cm]{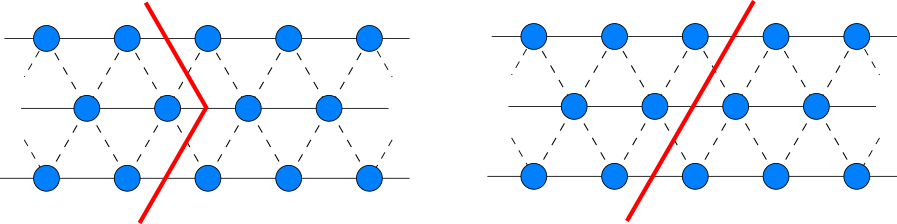}
\caption{Cut section of ATS model when calculating entanglement spectrum with iDMRG method. There are two types of cuts. One is V-shaped (shown in the left panel), and the other is straight (shown in the right panel).}
\label{fig:ats_model_infinite_cut}
\end{figure}

\begin{figure}[t]
\begin{center}
\includegraphics[width=8cm]{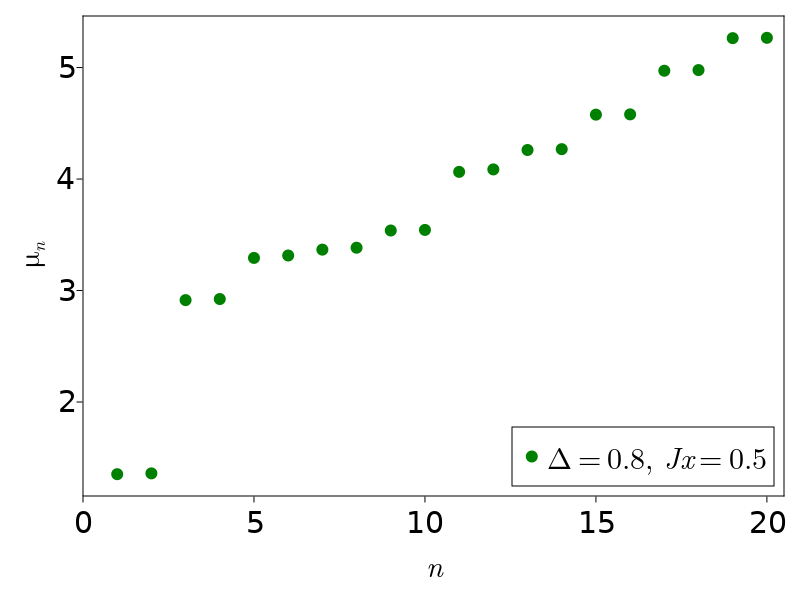}
\end{center}
\caption{The entanglement spectrum of ATS XXZ model for $\Delta=0.8$, $J_{\times}=0.5$ with V-shaped cutting. The horizontal axis is the numbering of the entanglement spectrum, and the vertical axis represents their values.}
\label{fig:entspec_inf_yabane}
\end{figure}

\begin{figure}[t]
\begin{center}
\includegraphics[width=8cm]{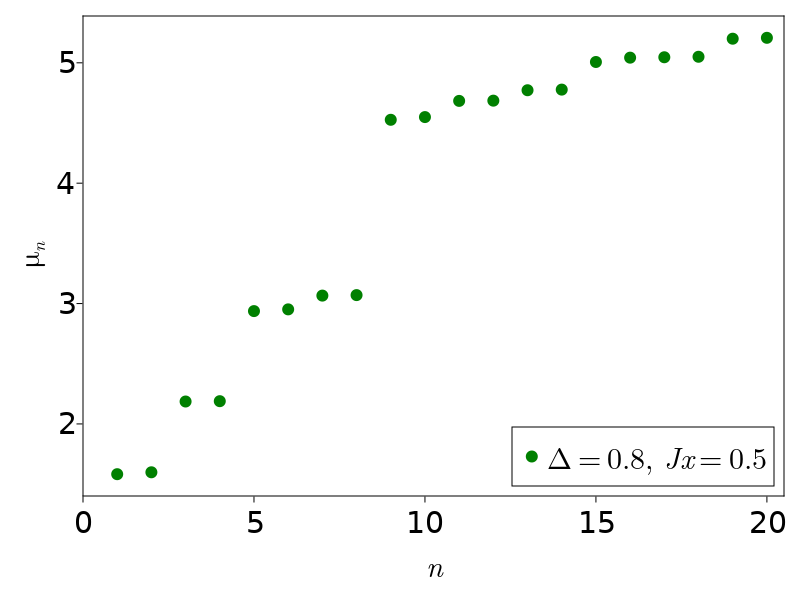}
\end{center}
\caption{The entanglement spectrum of ATS XXZ model for $\Delta=0.8$, $J_{\times}=0.5$ with parallelogram-type cutting. The horizontal axis is the numbering of the entanglement spectrum, and the vertical axis represents their values.}
\label{fig:entspec_inf_parallelo}
\end{figure}

To avoid this ambiguity about the finite-size cluster shape, we adopt the iDMRG method to further investigate the entanglement spectrum in the thermodynamic limit.
To calculate the entanglement spectrum with the iDMRG method, we can consider two types of cuts, as shown in Fig.~\ref{fig:ats_model_infinite_cut}. 
Note that both cuts respect the inversion symmetry in the infinite-size system in contrast to the finite-size cases.
Figure~\ref{fig:entspec_inf_yabane} shows the iDMRG result of the entanglement spectrum of the ATS XXZ model with the V-shaped cut,
and that for the straight cut is given in Fig.~\ref{fig:entspec_inf_parallelo}.
The both types of cuts lead to the even-fold degenerate entanglement spectra.
Moreover, the entanglement spectrum of the model with the straight cut resembles the one we obtained in the finite-size system with the parallel edges in Fig. \ref{fig:entspec_finite_parallelo}, as naively expected.
The even-fold degeneracy is entirely consistent with the simple physical picture that our gapless SPT state can be regarded essentially as a tensor product of the gapped Haldane state and gapless TLL. 
The even-fold degenerate entanglement spectrum implies the emergence of the spin-1/2 edge states akin to those in the spin-1 Haldane phase.
However, we cannot observe the edge state in the iDMRG calculations by construction or in the finite-size DMRG calculations because of the significant finite-size effects. 
We also note that no spontaneous symmetry breaking is found in the iDMRG calculations of the ground state of the ATS model.

Let us conclude this section.
The numerical result of the entanglement entropy implies that the ground state is the critical TLL state with $c=1$.
In addition, because of the even-fold degeneracy in the entanglement spectrum, the ground state simultaneously belongs to an SPT phase. 
Therefore, we can conclude that the ATS XXZ model with the parameters $\Delta=0.8$, $J_{\times}=0.5$ has the gapless SPT state as its ground state.
The following section gives a quantum field theoretical support to this claim.

\section{Effective field theory}\label{sec:Effective field theory}

This section develops a low-energy effective field theory of the ATS model.
This effective field theory explains how the geometrically frustrated interchain interaction enables the gapless SPT phase in the ATS model.
The effective field theory also uncovers that this gapless SPT phase qualifies as a symmetry-protected critical phase~\cite{furuya2017spc,yao2019spc}.

\subsection{Effective nearest-neighbor decoupling and ferromagnetic next-nearest-neighbor coupling}

To derive the effective field theory, we regard the interchain exchange $J_\times$ as a perturbation and set $\Delta=1$ for a while in this section.
We incorporate the intrachian easy-plane anisotropy into the effective field theory after taking the frustrated interchain interaction into account.
The ATS Heisenberg model's Hamiltonian consists of two parts.
\begin{align}
    H &= H_0 + V, 
    \label{H_Heisenberg}
    \\
    H_0 &= J \sum_j \sum_{n=1}^3 \vec S_{j,n} \cdot \vec S_{j+1,n},
    \label{H0_def} \\
    V &= J_\times \sum_j \sum_{n=1,3} \vec S_{j,2} \cdot (\vec S_{j,n} + \vec S_{j+1,n}).
    \label{V_def}
\end{align}
The unperturbed model with the Hamiltonian $H_0$ is made of three decoupled spin-1/2 Heisenberg chains.
Each Heisenberg chain has the TLL ground state~\cite{giamarchi_book}.

Let us deal with the single spin-1/2 Heisenberg chain based on a non-Abelian bosonization approach~\cite{gogolin_book}.
The spin operator $\vec S_{j,n}$ at low energies is represented in terms of two slowly varying fields,
\begin{align}
    \vec S_{j,n} &\approx \vec J_{n}(x)+ (-1)^j \vec N_n(x),
    \label{S2JN}
\end{align}
where $\vec J_n(x)$ and $\vec N_n(x)$ are the uniform and staggered parts of the spin operator.
As well as $\vec J_n$ and $\vec N_n$, a dimer operator,
\begin{align}
    (-1)^j \vec S_{j,n} \cdot \vec S_{j+1,n} &\approx \varepsilon_n(x),
\end{align}
plays fundamental roles in the bosonized theory.
The uniform part $\vec J_n$ is further split into two, $\vec J_n = \vec J_{R,n}+\vec J_{L,n}$. Here, $R$ and $L$ denote the right and left directions along which the boson field of the TLL propagates.
In particular, 
$J_{R,n}^z$ and $J_{L,n}^z$ are simply represented as
\begin{align}
    J_{R,n}^z &= \frac{1}{\sqrt{2\pi}}\partial_x\varphi_{R,n}, \quad J_{L,n}^z = \frac{1}{\sqrt{2\pi}} \partial_x \varphi_{L,n},
\end{align}
where $\varphi_{R,n}$ ($\varphi_{L,n}$) is the right-moving (left-moving) chiral boson of the TLL.
The unperturbed Hamiltonian is written as
\begin{align}
    H_0 &= \sum_{n=1}^3 \int dx \biggl[ \frac{2\pi v}{3} (\vec J_{R,n} \cdot \vec J_{R,n} + \vec J_{L,n} \cdot \vec J_{L,n})
    \notag \\
    &\qquad + \gamma_{\mathrm{bs}} \vec J_{R,n} \cdot \vec J_{L,n}
    \biggr],
    \label{H0_JJ}
\end{align}
where $v=\pi Ja_0/2$ is the velocity of the bosonic excitation of the TLL and $\gamma_{\mathrm{bs}}=O(J)>0$ represents the strength of the back scattering.
$a_0$ is the lattice spacing and hereafter set as unity unless otherwise stated.
The SU(2) spin rotational symmetry allows us to rewrite $\vec J_{R,n}\cdot \vec J_{R,n}=3(J_{R,n}^z)^2= 3(\partial_x\varphi_{R,n})^2/2\pi$ and $\vec J_{L,n}\cdot \vec J_{L,n} =3(\partial_x\varphi_{L,n})^2$.
The two chiral bosons $\varphi_{R,n}$ and $\varphi_{L,n}$ for each $n$ builds two (nonchiral) boson fields $\phi_n$ and $\theta_n$,
\begin{align}
    \phi_n &= \varphi_{L,n}  + \varphi_{R,n}, \\
    \theta_n &=  \varphi_{L,n} - \varphi_{R,n}.
\end{align}
The unperturbed Hamiltonian $H_0$ is thus given by
\begin{align}
    H_0 &= \sum_{n=1}^3 \int dx \biggl[ \frac v2 \{(\partial_x\theta_n)^2 + (\partial_x \phi_n)^2 \} + \gamma_{\mathrm{bs}} \vec J_{R,n} \cdot \vec J_{L,n}
    \biggr].
\end{align}
The back-scattering term is marginally irrelevant and mostly negligible in the TLL phase.
The back-scattering term affects neither the excitation spectrum nor the entanglement spectrum.
In the TLL phase,  the back-scattering term only adds quantitative corrections to physical quantities~\cite{eggert1994log_corr,Kobayashi2018_polarization,Furuya2019_polarization}.

Naively, the geometrically frustrated interchain interaction $V$ is composed of marginal or irrelevant interactions only.
\begin{align}
    V_{\rm naive} 
    &= \int dx \biggl[\gamma_J \vec J_2 \cdot (\vec J_1+\vec J_3) 
    +\gamma_{\rm tw} (\partial_x \vec N_2) \cdot (\vec N_1+ \vec N_3)
    \biggr],
    \label{V_naive}
\end{align}
with $\gamma_J=J_\times$ and $\gamma_{\rm tw}=J_\times$.
All the terms in Eq.~\eqref{V_naive} have the scaling dimension $2$ and  marginally irrelevant.
The naive representation \eqref{V_naive} is inaccurate because it misses relevant interactions generated in the course of the renormalization-group (RG) transformation~\cite{starykh2007triangular}.
We can show that the perturbation expansion about $V$ contains more interactions~\cite{starykh2007triangular,starykh2005checkerboard} (see also Appendix~\ref{app:ope}).
\begin{align}
    V_{\rm eff}
    &= \int dx \biggl[\gamma_J \vec J_2 \cdot (\vec J_1+\vec J_3) +\gamma_{\rm tw} (\partial_x \vec N_2) \cdot (\vec N_1+ \vec N_3)\biggr]
    \notag \\
    &\qquad + \int dx (\gamma'_J \vec J_1 \cdot \vec J_3+ \gamma_N \vec N_1 \cdot \vec N_3+\gamma_\varepsilon \varepsilon_1\varepsilon_3).
    \label{V_bosonized}
\end{align}
The additional terms $\vec J_1 \cdot \vec J_3$, $\vec N_1\cdot \vec N_3$, and $\varepsilon_1\varepsilon_3$  have the scaling dimension $2$ and $1$, and $1$ respectively.
The latter two thus can generate an excitation gap.
As we derive in Appendix.~\ref{app:ope}, the coupling constants $\gamma'_J$ and $\gamma_N$ are second order of $J_\times/J$:
\begin{align}
    \gamma'_J &= -\frac{J_\times^2}{2\pi^3J}, \\
    \gamma_N &= -\frac{J_\times^2}{4\pi^3J}.
    \label{gamma_N}
\end{align}
Note that $\gamma_N$ is the second order of $J_\times/J$.
A previous study~\cite{starykh2007triangular} concluded that $\gamma_N$ is $O(J(J_\times/J)^4)$.
We obtained qualitatively the same relevant interactions $\vec J_1\cdot \vec J_3$ and $\vec N_1\cdot \vec N_3$ as those derived in Ref.~\cite{starykh2007triangular} (see Eq. (5) therein). 
However, we found $\gamma_N$ is actually $O(J(J_\times/J)^2)$ (see Appendix~\ref{app:ope}).

The fourth-order correction to $V_{\rm eff}$ also includes another relevant interaction $\varepsilon_1\varepsilon_3$.
Since $\vec N_1\cdot \vec N_3$ and $\varepsilon_1\varepsilon_3$ have the same scaling dimension and $\gamma_N\gg \gamma_\varepsilon$, the ground state is governed by the strong-coupling limit of $\gamma_N$.

Note that $\gamma_N<0$ is ferromagnetic.
The frustrated interchain interaction $V\approx V_{\rm eff}$ develops the ferromagnetic interaction between the first and third legs.
The geometrical frustration makes the interchain interactions between the nearest-neighbor legs much weaker than those between the next-nearest-neighbor ones.
In particular, the interactions between the nearest-neighbor chains are marginally irrelevant in the RG sense.
Whereas the first and third legs are ferromagnetically coupled, the second leg is almost decoupled from the other legs.
Since the two-leg spin-$1/2$ ladder with a ferromagnetic interleg interaction has a SPT ground state that belongs to the spin-1 Haldane phase, the ground state of our system is approximately a product state of the TLL within the second leg and the gapped SPT state.
The present physical picture within the perturbation theory is expected to hold to some extent in an extended parameter region
and can provide a basis for understanding the numerical results in the previous section.

\section{Symmetry Protection}\label{sec:symmetryprotection}

In the previous sections, we showed that the ground state of the ATS XXZ model has simultaneously the gapless nature and the gapped SPT feature, which provides a naive definition of a gapless SPT state. 
However, a genuine gapless SPT state should be characterized by symmetries which protect the entire state including gapless and gapped sectors from symmetry-preserving perturbations.
In this section, we clarify the symmetry protection of our ground state and argue that it can indeed be understood as a genuine gapless SPT state but not just as a merely decoupled pair of a gapless TLL and a gapped SPT state.
We first point out that the gapless SPT phase hitherto investigated in this paper is a symmetry-protected critical (SPC) phase~\cite{furuya2017spc,yao2019spc}.
The SPC phase is characterized by ``ingappability''~\cite{cho2017lsm} under symmetries, that is, impossibility of opening an excitation gap with keeping the imposed symmetries.
According to Ref.~\cite{furuya2017spc}, the ground state of the spin-$1/2$ ATS model \eqref{H_Heisenberg} belongs to the SPC phase protected by the SU(2) spin-rotation symmetry, the one-site translation symmetry along the legs, and the emergent Lorentz symmetry~\cite{furuya2017spc}.

As our numerical results imply, we can relax the condition of the SU(2) spin-rotation symmetry to the $\mathrm{U(1)} \rtimes \mathbb{Z}_2$ symmetry without opening an excitation gap.
The $\mathrm{U(1)} \rtimes \mathbb{Z}_2$ group refers to the continuous spin rotation around the $z$ axis~\cite{metlitski2018_anomaly}.
This symmetry reduction is possible because, as we showed in the previous section, the criticality of the ATS Heisenberg model is basically attributed to the TLL of the Heisenberg antiferromagnetic model on the second leg.
The TLL is robust against the introduction of the easy-plane anisotropy.
However, the symmetries for the SPC state alone do not fully characterize the gapless SPT state.
We need an additional symmetry to protect the topological nontriviality of the gapless SPT phase besides the symmetry protected criticality.
Although one may naively expect that the symmetries protecting the spin-1 Haldane state will play the role, they are not enough as will be shown below.
Here, we provide a field theoretical argument on the symmetries required to simultaneously protect the ingappability and the topological nontriviality in our system.

\subsection{$D_2$ and time-reversal symmetric modification}

\begin{figure}[t]
\centering
\includegraphics[width=6cm]{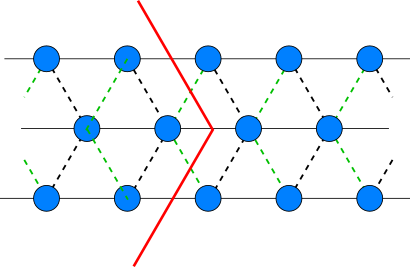}
\caption{The ATS XXZ model with the inversion symmetry-breaking interchain interaction, represented by the Hamiltonian \eqref{Ht_modification}. 
The green bonds represent the XXZ interaction with exchange coupling $(1-t)J_{\times}$.
The red line is the cut for the calculation of the entanglement spectrum.}
\label{fig:ats_model_breaksym_lattice}
\end{figure}

\begin{figure}[t]
\centering
\includegraphics[width=8cm]{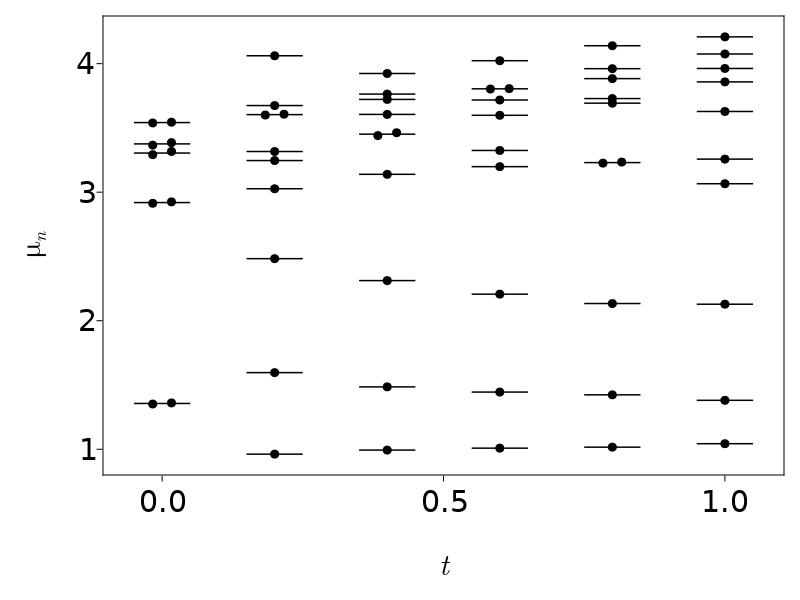}
\caption{Entanglement spectrum of the ATS XXZ model with additional unfrustrated interchain interaction \eqref{H'_def}. 
The horizontal axis is the value of $t$, and the vertical axis is the entanglement spectrum.}
\label{fig:ats_model_breaksymmetry_cut}
\end{figure}

First, to see that the symmetries for the spin-1 Haldane state do not protect our gapless SPT state, we consider the following modification of the ATS XXZ model (see Figure~\ref{fig:ats_model_breaksym_lattice}).
We introduce a real parameter $t\in [0,1]$ and modify the Hamiltonian \eqref{H_ATS} to
\begin{align}
    \mathcal H(t) = H + tH',
    \label{Ht_modification}
\end{align}
with an interchain interaction,
\begin{align}
    H' &= - J_\times \sum_j  \sum_{n=1,3} \vec S_{j,2}\cdot_\Delta \vec S_{j+1,n}.
    \label{H'_def}
\end{align}
$\mathcal H(t=0)$ gives the original spin-$1/2$ ATS XXZ Hamiltonian \eqref{H_ATS} but $\mathcal H(t=1)$ gives the Hamiltonian of a spin-$1/2$ \emph{unfrustrated} three-leg XXZ ladder:
\begin{align}
    \mathcal H(1)
    &= J \sum_j \sum_{n=1}^3 \vec S_{j,n} \cdot_\Delta \vec S_{j+1,n} + J_\times \sum_j \sum_{n=1,3} \vec S_{j,2} \cdot_\Delta \vec S_{j,n}.
    \label{H_unfrustrated}
\end{align}
For $J>0$ and $J_\times >0$, the unfrustrated spin ladder \eqref{H_unfrustrated} has the SPC phase for $\Delta\approx 1$.
However, this SPC phase is topologically trivial.
This can be clearly seen in degeneracy lifting of the entanglement spectrum.
Figure~\ref{fig:ats_model_breaksymmetry_cut} shows the iDMRG calculation results of the entanglement spectrum of model~\eqref{Ht_modification}. 
They indicate that the degeneracy of entanglement spectrum disappears as soon as we add the interaction (\ref{H'_def}) to the original ATS XXZ model.

The modification \eqref{Ht_modification} keeps the $D_2$ spin-rotation symmetry and the time-reversal symmetry.
Let us recall that the spin-1 Haldane phase is protected by one of the $D_2\cong \mathbb Z_2\times \mathbb Z_2$ spin-rotation symmetry, the time-reversal symmetry, and the bond-centered inversion symmetry.
Though the Hamiltonian \eqref{Ht_modification} keeps the $D_2$ spin-rotation and time-reversal symmetries for the entire range $0\le t \le 1$, this modification ruins the topological nontriviality.
The difference in symmetry protection from the spin-1 Haldane phase implies that we need to discuss the symmetry protection of topological nontriviality and the criticality at the same time to characterize the gapless SPT phase.

\subsection{Inversion symmetry}

Here, we show that an inversion symmetry along the leg direction protects the topological nontriviality of the gapless SPT phase.
The infinite-size ATS (Fig.~\ref{fig:ats_model}) with the preiodic boundary condition has a symmetry under the following inversion $\mathcal I$:
\begin{align}
    \mathcal I
    \begin{pmatrix}
    \vec S_{j,1} \\ \vec S_{j,2} \\ \vec S_{j,3}
    \end{pmatrix}
    \mathcal I^{-1} 
    &:= \begin{pmatrix}
    \vec S_{L+1-j,1} \\ \vec S_{L-j,2} \\ \vec S_{L+1-j,3}
    \end{pmatrix}.
\end{align}
This $\mathcal I$ inversion works as a site-centered inversion on the second leg but as a bond-centered one on the first and third legs.
In the perturbation theory in Sec.~\ref{sec:Effective field theory}, the subsystem made of the first and third legs effectively forms the spin-$1/2$ ladder with the ferromagnetic rung interaction whose ground state is the spin-1 Haldane state.
Within this subsystem, the $\mathcal I$ symmetry turns into the bond-centered inversion symmetry that protects the spin-1 Haldane phase.
Now we recall the modification \eqref{Ht_modification} that respect the $D_2$ spin-rotation and time-reversal symmetries.
The infinitesimal $t\not=0$ breaks the even-fold degeneracy of the entanglement spectrum because the added interaction \eqref{H'_def} breaks the $\mathcal I$ symmetry.
Therefore, the $\mathcal I$ symmetry imposes $t=0$ in the Hamiltonian \eqref{Ht_modification}.

In the gapless SPT phase of the ATS XXZ model, the topological nontriviality is protected by the inversion symmetry and by neither the $D_2$ spin-rotation symmetry nor time-reversal symmetry, different from the gapped SPT phase (the spin-1 Haldane phase).
The geometrical frustration of interchain interactions make the $\mathcal I$ inversion symmetry special and different from the $D_2$ and time-reversal symmetries.
As we saw, the ground state becomes topologically trivial as soon as the geometrical frustration is resolved (Fig.~\ref{fig:ats_model_breaksymmetry_cut}).

Furthermore, we point out that the $\mathcal I$ symmetry also protects the effective decoupling of the total system into the gapless TLL sector and gapped SPT sector at low energy.
In terms of the effective field theory, the geometrical frustration forbids the relevant interchain interactions in the sense of the renormalization group.
A complete list of such relevant interchain interactions is available.
The list is given by
\begin{align}
    \{\vec N_2 \cdot \vec N_n, \, \varepsilon_2\varepsilon_n, \, N_2^a\varepsilon_n, \, \varepsilon_2N_n^a\},
    \label{list_relevant}
\end{align}
for $n=1,3$ and $a=x,y,z$.
The last two interactions $N_2^a\varepsilon_n$ and $\varepsilon_2N_n^a$ are forbidden by the $\mathrm{U(1)}$ spin-rotation symmetry.
The $\mathcal{I}$ inversion symmetry excludes the remaining two interactions, $\vec N_2\cdot \vec N_n$ and $\varepsilon_2\varepsilon_n$ for $n=1,3$, from the effective field theoretical Hamiltonian.

We can directly show that the $\mathcal I$ symmetry forbids both $\vec N_2 \cdot \vec N_n$ and $\varepsilon_2\varepsilon_n$.
The interchain interaction \eqref{H'_def} introduces relevant interactions such as
\begin{align}
    V' &=  \gamma' \int dx \,  \vec N_2 \cdot (\vec N_1 + \vec N_3),
    \label{H'_bosonized}
\end{align}
with $\gamma' \propto J_\times$.
The relevant interaction \eqref{H'_bosonized} violates the gapless second leg from the topological subsystem of the first and third legs.
The $\mathcal I$ symmetry indeed forbids the interaction \eqref{H'_bosonized} to enter into the Hamiltonian.
$\mathcal I$ acts on $\vec N_n$ and $\varepsilon_n$ as
\begin{align}
    \mathcal I
    \begin{pmatrix}
    \vec N_{1}(x) \\ \vec N_{2}(x) \\ \vec N_{3}(x)
    \end{pmatrix}
    \mathcal I^{-1} 
    &= \begin{pmatrix}
    -\vec N_{1}(-x) \\ \vec N_{2}(-x) \\ -\vec N_{3}(-x)
    \end{pmatrix}
    \label{inversion_N}
\end{align}
and
\begin{align}
    \mathcal I
    \begin{pmatrix}
    \varepsilon_{1}(x) \\ \varepsilon_{2}(x) \\ \varepsilon_{3}(x)
    \end{pmatrix}
    \mathcal I^{-1} 
    &= \begin{pmatrix}
    \varepsilon_{1}(-x) \\ -\varepsilon_{2}(-x) \\ \varepsilon_{3}(-x)
    \end{pmatrix}.
    \label{inversion_epsilon}
\end{align}
Equations~\eqref{inversion_N} and \eqref{inversion_epsilon} indicate that the $\mathcal I$ symmetry forbids both $\vec N_2 \cdot \vec N_n$ and $\varepsilon_2\varepsilon_n$ for $n=1$ and $3$.

Therefore, we conclude that the $\mathrm{U(1)}$ spin symmetry, the translation symmetry, and the $\mathcal I$ inversion symmetry simultaneously protect the ingappability and the topological nontriviality of the gapless SPT phase of the ATS XXZ model \eqref{H_ATS}.
The former two symmetries protect the ingappability and the last one protects both the effective decoupling and the topological nontriviality.
The symmetry protection clearly distinguishes the present gapless SPT state from an independent pair of a gapless TLL and a gapped SPT state.

\section{Summary and discussions}
In this work, we introduced the ATS XXZ model and showed that this model exhibits the gapless SPT state. From our DMRG calculations of the entanglement entropy, we found that the ground state is the critical TLL with the central charge $c=1$. 
We also calculated the entanglement spectrum with the finite-size DMRG and iDMRG methods.
We confirmed the evenfold degeneracy of the entanglement spectrum.

We also analyzed the ATS XXZ model with the quantum field theory. 
The geometrically frustrated inter-chain interaction effectively decouples the second leg from the first and third legs.
Nevertheless, the second leg mediates the ferromagnetic interaction between the first and third legs.
At low energies, the effectively decoupled second leg behaves as the TLL, whereas the spin ladder formed by the first and third chains behaves as the spin-1 Haldane state. 
As a whole, the ATS XXZ model forms the gapless SPT phase. 
This gapless SPT phase is protected by the U(1) spin-rotation, translation, \emph{and} the $\mathcal I$ inversion symmetry.

In this work, we considered the parameter region with small $J_\times/J$ in order to compare the numerical result with the effective field theory. 
We have numerically checked that our gapless SPT phase is stable to changes of $J_\times/J$ and $\Delta$ to a certain extent. Showing a detailed $J_\times-\Delta$ phase diagram is beyond the scope of this paper and left for a future study~\cite{Hidakapreparation}.
It will be interesting to investigate the ATS XXZ model in a wide parameter region, in particular, with large $J_\times/J$.
When $J=0$, the ATS XXZ model is reduced to an experimentally feasible model, a spin-1/2 diamond chain~\cite{Okamoto2003_diamond,kikuchi2005_azurite}.
Since the diamond chain does not show the gapless SPT phase, we expect that there will be a phase transition from the gapless SPT phase as we increase $J_\times/J$.

The ATS XXZ model is simple but turns out to be highly nontrivial.
It hosts the gapless SPT phase where the ground state is approximately a product state of the critical TLL and the spin-1 Haldane state, in which the effective decoupling is constrained by the inversion symmetry.
At the same time, the simple structure of the ATS XXZ model helped us to foster a better understanding of the symmetry protection of the gapless SPT phase.
We believe that the ATS XXZ model plays a fundamental role in future studies on the classification of gapless phases.

It is noteworthy briefly mentioning the feasibility of the ATS model.
Natural minerals Cu$_3$(OH)$_4$MO$_4$ (M$=$ S, Mo) are known as spin-1/2 triple-chain magnets composed of three spin-1/2 chains with zigzag interchain interactions just like our ATS model~\cite{Vilminot2003_antlerite,Okubo2009_antlerite,Fujii2009_antlerite,Hara2011_antlerite,Fujii2013_antlerite,Fujii2015_szenicsite,Lebernegg2017_szenicsite}.
However, unfortunately, these compounds consist of complex exchange interactions that break the one-site translation symmetry.
Still, we hope that our study will stimulate further experimental studies about such triple-chain systems.

\begin{acknowledgments}
The authors are grateful to Shunsuke Furukawa, Katsuhiro Morita, Kiyomi Okamoto, Masaki Oshikawa, and Hiroshi Ueda for fruitful discussions.
This work was supported by a Grant-in-Aid for Scientific Research on Innovative Areas “Quantum Liquid Crystals,” Grant No. JP19H05825 (for S.C.F.), JSPS Grants-in-Aid for Transformative Research Areas (A) ``Extreme Universe,'' Grants No. JP21H05191 and No. 21H05182  (for S.C.F.), and
JSPS KAKENHI Grants No. JP20K03769 (for S.C.F), No. JP21K03465 (for S.C.F.), No. JP17K14333 (for Y.T.), No. JP22K03513 (for Y.T.), and
No. JP21J20523 (for A.U.).
\end{acknowledgments}

\appendix

\section{String order}
\label{app:string}
\begin{figure}[t]
\centering
\includegraphics[width=8cm]{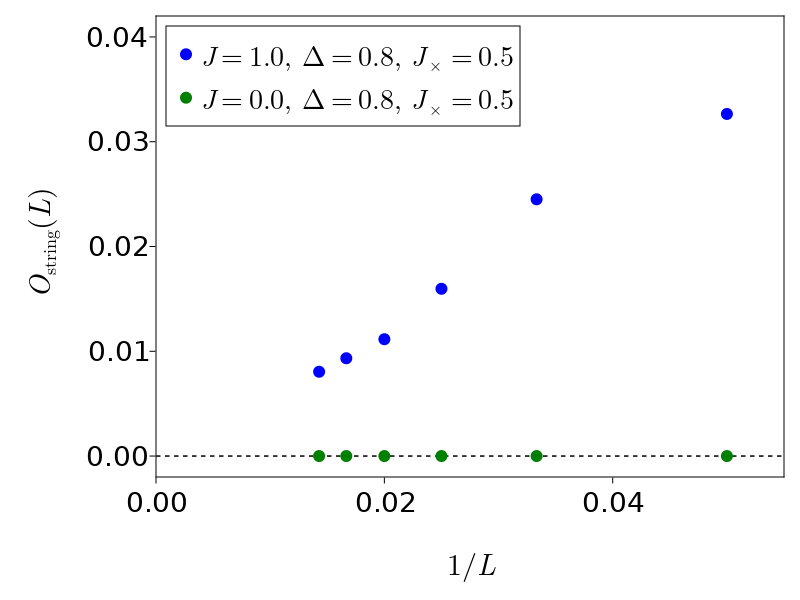}
\caption{$L$ dependence of string order $\mathcal{O}_{\mathrm{string}}(L/4,3L/4;L)$. When $J=1.0$, $\Delta=0.8$, and $J_{\times}=0.5$, the small string order emerges (blue circles). For $J=0.0$, the string order becomes completely zero (green circles).}
\label{fig:string}
\end{figure}

In this section, we briefly discuss a string order of the ATS model.
Let us define the string order between $i$-th and $j$-th column of the ATS model with length $L$ as follows:
\begin{align}
&\mathcal{O}_{\mathrm{string}}(i,j;L)\nonumber\\&=-\Braket{\psi_L|T_{i}^z\mathrm{exp}\left[i\pi\sum_{k=i+1}^{j-1}T_{k}^z\right]T_{j}^z|\psi_L},
\end{align}
We have introduced $T_{i}^z=S_{i,1}^{z}+S_{i,3}^{z}$, since the ground state contains the ferromagnetic ladder of the first and third legs as discussed in Sec.~\ref{sec:Effective field theory}. The accurate definition of string order (in infinite length) is $\mathcal{O}_{\mathrm{string}}=\lim_{|i-j|\to\infty}\lim_{L\to\infty}\mathcal{O}_{\mathrm{string}}(i,j;L)$. However, instead of calculating the infinite-size string order, we calculated $\mathcal{O}_{\mathrm{string}}(L/4,3L/4;L)$ with the finite-size DMRG method and investigate its size dependence. 
The numerical results are shown in Fig.~\ref{fig:string}. 
The effective ferromagnetic interchain interaction between the first and third chains is expected to be small.
The string order $\mathcal O_{\rm string}(L/4,3L/4;L)$ will also become small. 
We numerically confirmed that the string order takes a finite value in the finite-size system. 
In contrast, the string order for $J=0$ is completely zero irrespective of the system size.
For $J=1$, $J_\times=0.5$, and $\Delta=0.8$, the string order is decreasing as the system size increases.
Unfortunately, extrapolation to $L\to\infty$ is difficult and we cannot draw a definite conclusion for the string order parameter in the thermodynamic limit.
On the other hand, our characterization of the topological state in the main text is based on the entanglement spectrum, which turned out to be more robust to the finite-size effects than the string order parameter.

\section{Perturbative analyses of interchain interactions}
\label{app:ope}

This section supplements the perturbative expression \eqref{V_bosonized} of the interchain interaction.
Our arguments are similar to those in Refs.~\cite{starykh2007triangular,starykh2005checkerboard,furukawa2012xxz}.
The present effective field theory stands on three copies of the level-1 SU(2) Wess-Zumino-Witten (WZW) theories weakly coupled to each other.
Using the operator product expansion of the level-1 SU(2) WZW theory, we derive an effective Euclidean action of the low-energy effective field theory.

\subsection{Operator product expansions}

Let us introduce the operator product expansion of $\vec J_{R,n}$, $\vec J_{L,n}$, $\vec N_n$, and their derivatives.
Since the operator product expansion works for operators that share the leg index $n$, we omit the index $n$ in this subsection for simplicity.
We can represent $\vec J_R$, $\vec J_L$, $\vec N$, and $\varepsilon$ in terms of free chiral Dirac fermions $\psi_{R,s}$ and $\psi_{L,s}$ with spin $s=\uparrow, \downarrow$:
\begin{align}
    \vec J_R &= :\psi_{R,s}^\dag \frac{\vec \sigma_{s,s'}}{2}\psi_{R,s'}:,
    \label{JR_fermion_def} \\
    \vec J_L &= :\psi_{L,s}^\dag \frac{\vec \sigma_{s,s'}}{2}\psi_{L,s'}:,
    \label{JL_fermion_def} \\
    \vec N &= :\psi_{R,s}^\dag \frac{\vec \sigma_{s,s'}}{2}\psi_{L,s'}: + :\psi_{L,s}^\dag \frac{\vec \sigma_{s,s'}}{2}\psi_{R,s'}:,
    \label{N_fermion_def} \\
    \varepsilon &= \frac i2 (:\psi_{R,s}^\dag \psi_{L,s}: - :\psi_{L,s}^\dag \psi_{R,s}:),
    \label{epsilon_fermion_def}
\end{align}
where $\vec \sigma=(\sigma^x\, \sigma^y\,\sigma^z)^\top$ is the set of the Pauli matrices and $:\cdot:$ denotes the normal ordering.
These chiral fermions satisfy the following correlation functions at zero temperature,
\begin{align}
    \braket{\mathcal T\psi_{R,s}(x,\tau)\psi_{R,s'}^\dag(0,0)}
    &=  \frac{\delta_{s,s'}}{2\pi[v\tau - ix + a_0 \sigma(\tau)]}, 
    \label{psi_R_corr} \\
    \braket{\mathcal T\psi_{L,s}(x,\tau)\psi_{L,s'}^\dag(0,0)}
    &=  \frac{\delta_{s,s'}}{2\pi[v\tau + ix + a_0 \sigma(\tau)]}, 
    \label{psi_L_corr}
\end{align}
where $\tau$ is the imaginary time, $\mathcal T$ denotes the imaginary-time ordering and $\sigma(\tau)$ denotes the sign of $\tau$,
\begin{align}
    \sigma(\tau) &= \left\{
    \begin{array}{ccc}
    1 && (\tau>0) \\
    0 && (\tau=0) \\
    -1 && (\tau<0)
    \end{array}
    \right..
\end{align}
Here, we explicitly introduced the lattice spacing $a_0$ that was set as unity in the main text.

It is convenient to introduce two complex coordinates,
$z_L = v\tau+ix$ and $z_R=v\tau-ix$ that denote the space-(imaginary)time position of the right-moving and left-moving particles.
Note that $\vec J_R(z_R)$ and $\vec J_L(z_L)$ are independent of the coordinate that corresponds to the opposite chirality.
Nonchiral operators $\vec N(z_R,z_L)$ and $\varepsilon(z_R,z_L)$ depend on both coordinates.
Hereafter, we omit the normal and imaginary-time orderings following the convention.

The Wick's theorem leads to the following operator product expansions~\cite{difrancesco_yellow,furukawa2012xxz,Metavitsiadis2017_OPE}.
\begin{widetext}
\begin{align}
    J_R^a(z_R)J_R^b(0) &= \frac{\delta^{ab}}{8\pi^2[z_R+a_0\sigma(\tau)]^2} 
    + \frac{i\epsilon^{abc}}{2\pi[z_R+a_0\sigma(\tau)]}J_R^c(0), \\
    J_L^a(z_L)J_L^b(0) &= \frac{\delta^{ab}}{8\pi^2[z_L+a_0\sigma(\tau)]^2} 
    + \frac{i\epsilon^{abc}}{2\pi[z_L+a_0\sigma(\tau)]}J_L^c(0), \\
    J_R^a(z_R) N^b(0,0) &= \frac{1}{4\pi[z_R+a_0\sigma(\tau)]}\biggl( i\epsilon^{abc}N^c(0,0)-i\delta^{ab}\varepsilon(0,0) \biggr),
    \label{JRN_OPE}
    \\
    J_L^a(z_R) N^b(0,0) &= \frac{1}{4\pi[z_L+a_0\sigma(\tau)]}\biggl( i\epsilon^{abc}N^c(0,0)+i\delta^{ab}\varepsilon(0,0) \biggr),
    \label{JLN_OPE}
    \\
    J_R^a(z_R)\partial_x N^b(0,0) &= \frac{1}{4\pi[z_R+a_0\sigma(\tau)]^2} \biggl( \epsilon^{abc} N^c(0,0)-\delta^{ab}\varepsilon(0,0) \biggr)+ \cdots,
    \label{JRdN_OPE}
    \\
    J_L^a(z_L)\partial_x N^b(0,0) &= \frac{1}{4\pi[z_L+a_0\sigma(\tau)]^2} \biggl( -\epsilon^{abc} N^c(0,0)-\delta^{ab}\varepsilon(0,0) \biggr)+ \cdots,
    \label{JLdN_OPE}
    \\
    N^a(z_R,z_L) N^b(0,0)
    &= \sqrt{[z_R+a_0\sigma(\tau)][z_L+a_0\sigma(\tau)]}\biggl[
    \frac{\delta^{ab}}{4\pi^2[z_R+a_0\sigma(\tau)][z_L+a_0\sigma(\tau)]} 
    \notag \\
    &\qquad 
    + i\epsilon^{abc} \biggl( \frac{J_R^c(0)}{2\pi[z_L+a_0\sigma(\tau)]} + \frac{J_L^c(0)}{2\pi[z_R+a_0\sigma(\tau)]} \biggr) + \cdots \biggr], \\
    N^a(z_R,z_L) \partial_x N^b(0,0)
    &= \sqrt{[z_R+a_0\sigma(\tau)][z_L+a_0\sigma(\tau)]}\biggl[
    \frac{-i\delta^{ab}}{8\pi^2[z_R+a_0\sigma(\tau)]^2[z_L+a_0\sigma(\tau)]} 
    \notag \\
    &\qquad +  \frac{i\delta^{ab}}{8\pi^2[z_R+a_0\sigma(\tau)][z_L+a_0\sigma(\tau)]^2}
    + i\epsilon^{abc} \biggl( \frac{iJ_R^c(0)}{4\pi[z_L+a_0\sigma(\tau)]^2} + \frac{-iJ_L^c(0)}{4\pi[z_R+a_0\sigma(\tau)]^2} \biggr)
    \notag \\
    &\qquad +i\epsilon^{abc}\frac{iJ_R^c(0)-iJ_L^c(0)}{4\pi[z_R+a_0\sigma(\tau)][z_L+a_0\sigma(\tau)]}
    + \cdots\biggr], \\
    [\partial_xN^a(z_R,z_L)] [\partial_xN^b(0,0)]
    &= \sqrt{[z_R+a_0\sigma(\tau)][z_L+a_0\sigma(\tau)]}\biggl[\frac{3\delta^{ab}}{16\pi^2[z_R+a_0\sigma(\tau)]^3[z_L+a_0\sigma(\tau)]} 
    \notag \\
    &\qquad +\frac{-\delta^{ab}}{8\pi^2[z_R+a_0\sigma(\tau)]^2[z_L+a_0\sigma(\tau)]^2} 
    + \frac{3\delta^{ab}}{16\pi^2[z_R+a_0\sigma(\tau)][z_L+a_0\sigma(\tau)]^3} 
    \notag \\
    &\qquad + i\epsilon^{abc}\biggl(\frac{3J_R^c(0)}{8\pi[z_L+a_0\sigma(\tau)]^3} + \frac{3J_L^c(0)}{8\pi[z_R+a_0\sigma(\tau)]^3} \biggr) 
    \notag \\
    &\qquad + i\epsilon^{abc} \biggl(\frac{J_R^c(0)-J_L^c(0)}{8\pi[z_R+a_0\sigma(\tau)][z_L+a_0\sigma(\tau)]^2} + \frac{-J_R^c(0)+J_L^c(0)}{[z_R+a_0\sigma(\tau)]^2[z_L+a_0\sigma(\tau)]}
    \biggr)
    + \cdots\biggr],
\end{align}
\end{widetext}
where $\delta^{ab}$ is the Kronecker's delta, $\epsilon^{abc}$ is the completely antisymmetric tensor with $\epsilon^{xyz}=1$, and the terms denoted by $\cdots$ are omitted above because they are irrelevant in the RG sense.
We emphasize two points about these operator product expansions.
First, the operator product expansions hold when $|z_R|\ll 1$ and $|z_L|\ll 1$.
Second, the right hand sides contain much more relevant operators than the left hand sides [e.g., Eqs.~\eqref{JRN_OPE} and \eqref{JRdN_OPE}].

\subsection{Effective Euclidean action}

Let $\mathcal S$ be the Euclidean action of the low-energy effective field theory.
We can represent $\mathcal S$ as an imaginary-time integral of the following Hamiltonian.
\begin{align}
    \mathcal S &= \mathcal S_0 + \mathcal S_\times,
    \label{S_full_def} \\
    \mathcal S_0 &= \int_{-\infty}^\infty d\tau \, \mathcal H_0(\tau), 
    \label{S0_def} \\
    \mathcal S_\times &= \int_{-\infty}^\infty d\tau\, V_{\rm naive}(\tau).
    \label{Sx_def}
\end{align}
The naive expression \eqref{V_naive} is correct here because we obtain it by replacing the spin operator with $\vec J_n$ and $\vec N$ [Eq.~\eqref{S2JN}].
The full partition function $Z$ and the unperturbed one $Z_0$ are symbolically represented in terms of path integrals as,
\begin{align}
    Z &= \int \prod_{n=1}^3 \mathcal D \vec J_n \mathcal D \vec N_n \exp(-\mathcal S),
    \label{Z_path_int} \\
    Z_0 &= \int \prod_{n=1}^3 \mathcal D \vec J_n \mathcal D \vec N_n \exp(-\mathcal S_0).
    \label{Z0_path_int}
\end{align}
We perform the perturbation expansion,
\begin{align}
    \exp(-\mathcal S_\times)=1-\mathcal S_\times + \frac 12 \mathcal S_\times^2 + \cdots,
\end{align}
up to the second order of $V_{\rm naive}$.
The second-order term $\mathcal S_\times^2=\int_{-\infty}^\infty d\tau' \int_{-\infty}^\infty d\tau V_{\rm naive} (\tau) V_{\rm naive}(\tau')$ contain many nonlocal interactions such as $[\vec J_2(x,\tau) \cdot \vec J_1(x,\tau)][\vec J_2(x',\tau')\cdot \vec J_3(x',\tau')]$.
We already saw that RG relevant interactions emerge when the operator product expansion works, namely for $x\approx x'$ and $\tau\approx \tau'$. 
Following Ref.~\cite{starykh2005checkerboard}, we keep the relevant interaction with $x= x'$ and $\tau\approx \tau'$ in $V_{\rm naive}(\tau)V_{\rm naive}(\tau')$.
The operator product expansions given above lead to
\begin{align}
    &[\vec J_2(x,\tau)\cdot \vec J_1(x,\tau)][\vec J_2(x,\tau')\cdot \vec J_3(x,\tau')]
    \notag \\
    &= \frac{1}{4\pi^2[v(\tau-\tau')+a_0\sigma(\tau-\tau')]^2} \vec J_1(x,\tau') \cdot \vec J_3(x,\tau') + \cdots, \\
    &[\partial_x \vec N_2(x,\tau) \cdot \vec N_1(x,\tau)][\partial_x \vec N_2(x,\tau) \cdot \vec N_3(x,\tau')]
    \notag \\
    &= \frac{1}{4\pi^2|v(\tau-\tau')+a_0\sigma(\tau-\tau')|^3} \vec N_1(x,\tau') \cdot \vec N_3(x,\tau') 
    \notag \\
    &\qquad + \cdots.
\end{align}
$\mathcal S_\times^2$ generates interactions between the first and third legs through operator product expansions.
We thus obtain
\begin{align}
    \frac 12\mathcal S_\times^2
    &\approx \int_{-\infty}^\infty d\tau \int dx \biggl[\frac{\gamma_J^2}{4\pi^2va_0} \vec J_1 \cdot \vec J_3 + \frac{\gamma_{\rm tw}^2}{8\pi^2va_0^2} \vec N_1 \cdot \vec N_3 
    \biggr].
\end{align}
These interactions turns into the effective \emph{ferromagnetic} interaction between the first and third legs because we may approximate $\exp(-\mathcal S_\times)$ as
\begin{align}
    \exp(-\mathcal S_\times)
    &\approx 1 - \mathcal S_\times + \frac 12 (\mathcal S_\times)^2 \notag \\
    &\approx 1- \mathcal S'_\times
    \notag \\
    &\approx \exp(-\mathcal S'_\times),
\end{align}
where $\mathcal S'_\times$ is given by
\begin{align}
    \mathcal S'_\times 
    &\approx \int d\tau \, V_{\rm naive}(\tau)
    \notag \\
    &\qquad - \int d\tau dx \biggl[\frac{\gamma_J^2}{4\pi^2v} \vec J_1 \cdot \vec J_3 + \frac{\gamma_{\rm tw}^2}{8\pi^2 v} \vec N_1 \cdot \vec N_3
    \biggr].
\end{align}
Note that we obtained qualitatively the same result as Ref.~\cite{starykh2007triangular} (see Eq. (5) therein).
However, we comment that the coupling constant $\gamma_{\rm tw}^2/8\pi^2v \sim J_\times^2/J$ is the second-order of $J_\times^2$ while it is fourth in Ref.~\cite{starykh2007triangular}.

%
\end{document}